\documentstyle[prl,aps,multicol,eqsecnum,epsfig]{revtex}

\begin{document}


\def\nid{\noindent}
\def\half{\mbox{\small $\frac{1}{2}$}}
\def\quarter{\mbox{\small $\frac{1}{4}$}}
\def\sfrac#1#2{\mbox{\small $\frac{#1}{#2}$}}
\def\beq{\begin{equation}}
\def\eeq{\end{equation}}
\def\eeql#1{\label{#1} \end{equation}}
\def\bea{\begin{eqnarray}}
\def\eea{\end{eqnarray}}
\def\eeal#1{\label{#1} \end{eqnarray}}
\def\sech{\mathop{\rm sech}\nolimits}
\def\scalefig#1{\def\epsfsize##1##2{#1##1}}
\def\vec{\bbox}
\hyphenation{anti-nodal}


\title{SUSY Transformations for Quasinormal Modes
of Open Systems}

\author{P.~T.~Leung${}^{(1)}$, 
Alec~Maassen van den Brink${}^{(1,2)}$,
W.~M.~Suen${}^{(1,3)}$, \\
C.~W.~Wong${}^{(1)}$
and K.~Young${}^{(1)}$}

\address{${}^{(1)}$Physics Department,
The Chinese University of Hong Kong, 
Hong Kong, China}

\address{${}^{(2)}$Department of Physics and Astronomy, 
SUNY Stony Brook,\\
Stony Brook, NY 11794-3800, USA}

\address{${}^{(3)}$McDonnell Center for Space Sciences, Department of Physics,
Washington University, \\
St.~Louis, MO 63130-4899, USA}

\date{\today}

\maketitle



%

\begin{abstract}

Supersymmetry (SUSY) in quantum mechanics is extended
from square-integrable states to those satisfying
the outgoing-wave boundary condition, in a Klein--Gordon formulation.
This boundary condition allows both the usual normal modes and
quasinormal modes with complex eigenvalues $\omega$.  
The simple generalization leads to three features:
the counting of eigenstates under SUSY becomes more systematic;
the linear-space structure of outgoing waves (nontrivially
different from the usual Hilbert space of square-integrable states)
is preserved by SUSY; and multiple states at the same frequency
(not allowed for normal modes) are also preserved.
The existence or otherwise of SUSY partners is furthermore 
relevant to the question of inversion: 
are open systems uniquely determined by their complex
outgoing-wave spectra?  

\end{abstract}



\begin{multicols}{2}




%

\section{Introduction}
\label{sect:intro}

\subsection{Outline}
\label{subsect:outline}


Supersymmetry (SUSY) in quantum field theory~\cite{susyqft,weinfar}
relates bosons to fermions.  Its analog in 
quantum mechanics~\cite{witten,susyqm} is interesting in
its own right, and relates two
(typically one-dimensional) Hamiltonians $H$ and ${\tilde H}$
with the same spectrum of normal modes (NMs), or the
same spectrum apart from one state.  The classical limit~\cite{susycl} 
relates a one-parameter family of Hamiltonians $H(\xi)$.

This paper generalizes SUSY in quantum mechanics
to open systems and in particular to their
quasinormal modes (QNMs), which are eigenfunctions satisfying
the outgoing-wave condition (OWC) at infinity.
The generalization itself is straightforward, 
and we focus on the novel features that ensue.
After a brief introduction to open systems and QNMs
in the rest of this Section, this paper presents three main
results.

First, in the conventional discussion for the
Schr\"odinger equation, spectrum preservation 
can be conveniently expressed
in terms of the number of NMs $n(E)$ for $H$ and ${\tilde n}(E)$
for ${\tilde H}$ at the same energy $E$ .  Namely that
$n(E) = {\tilde n}(E)$ for all real $E$ except a privileged
value $E_0$ (usually chosen to be $E_0=0$).
At this value, the difference is given by the Witten index~\cite{witten}
 
\beq
\Delta(E_0) = n(E_0) - {\tilde n}(E_0) \quad ,
\eeql{eq:index1}

\nid
which can be $+1$, $0$, or $-1$.  For open systems with the OWC,
time-reversal invariance is broken,
and it is appropriate to consider a Klein--Gordon equation
(KGE) instead --- in effect
replacing $\partial_t \mapsto \partial_t^2$
or $E \mapsto \omega^2$ in the time-independent
equation, and distinguishing $+\omega$
from $-\omega$ (since reversing the two converts an
outgoing to an incoming wave)~\cite{conv}.
Section~\ref{sect:form} shows that the equality
of spectrum for the KGE extends to the complex $\omega$-plane
except at the two privileged frequencies $\pm \Omega = \pm \sqrt{E_0}$,
namely

\beq
n(\omega) = {\tilde n}(\omega) 
\quad, \quad \omega \ne \pm \Omega \quad.
\eeql{eq:index2a}

\nid
At $\pm \Omega$, one needs to consider

\beq
\Delta(\pm\Omega) = n(\pm\Omega) - {\tilde n}(\pm\Omega) \quad .
\eeql{eq:index2}

\nid
These are again $+1$, $0$, or $-1$, but with
$\Delta(\Omega) = - \Delta(-\Omega)$ determined by
the asymptotic behavior of the SUSY generator, to be defined below.
In other words, under $H \mapsto {\tilde H}$,
if a state is removed (added) at $\Omega$, then a state
is added (removed) at $-\Omega$.
This relationship, which also applies to conservative systems and
NMs provided we take the KGE point of view, sharpens the 
information provided by (\ref{eq:index1}).

Secondly, SUSY preserves norms and inner products.
However, for outgoing waves the usual
norms and inner products are not useful.
For example, an outgoing wave of frequency $\omega$
goes as $\exp[i\omega (|x|-t)]$ at spatial infinity.
With $\mbox{Im } \omega < 0$ for QNMs
(see Section~\ref{subsect:qnm}), the 
exponential growth in $|x|$ renders the wavefunction not
normalizable in the usual sense.  A generalized
norm for QNMs was first introduced by Zeldovich~\cite{zel} many years
ago, and shown to be useful for time-independent
perturbation theory (of the {\em complex\/} eigenvalues)~\cite{pert1}.
An associated generalized inner product can also be defined~\cite{twocomp}.
The time-evolution operator turns out to be
symmetric under this product (the analog of self-adjoint).
Section~\ref{sect:orth} shows that SUSY preserves these generalized norms and
inner products --- a pleasant surprise, 
since their construction is totally unrelated to SUSY.

Thirdly, for non-conservative systems, there is no guarantee that the Hamiltonian can
be completely diagonalized; in general the best that one can do is to
decompose it into Jordan blocks (JBs)~\cite{jb0}.  Each block $j$, say of size $M_j \times M_j$, 
is associated with an eigenfrequency $\omega_j$, with $M_j=1$
being the usual case of a QNM.  Section~\ref{sect:jb} shows that
SUSY preserves the JB structure: except for
$\omega_j = \pm \Omega$, a block of size $M_j$
maps to a block also of size $M_j$ at the same
frequency $\omega_j$.
In fact, if we generalize the definition of $n(\omega_j)$ to be the order
$M_j$ of the block, then the relationship 
between $n(\omega)$ and ${\tilde n}(\omega)$ 
[cf.~(\ref{eq:index2a}) and (\ref{eq:index2})]
remains valid even for $n , {\tilde n} > 1$.

Examples are given in Section~\ref{sect:ex} and a discussion
is presented in Section~\ref{sect:disc}, including a
sketch of some issues for potentials with tails --- which can lead
to situations with {\em negative\/} $n$ and ${\tilde n}$
that are nevertheless accommodated in the same formalism. 

\subsection{Quasinormal modes}
\label{subsect:qnm}

In open systems, waves are not confined, but can escape:
acoustic waves from a musical instrument, electromagnetic waves
from a laser, and linearized gravitational waves
from a Schwarzschild background (to infinity and into the horizon).  
These systems are often
described (e.g., in the case of gravitational waves~\cite{chand})
by the KGE

\beq
\left[ \partial_t^2 - \partial_x^2 + V(x) \right]
\phi(x,t) = 0 \quad ,
\eeql{eq:kgt}

\nid
or (e.g., in the case of optics~\cite{lamb})
by the closely related~\cite{twocomp} wave equation

\beq
\left[ \rho(x) \partial_t^2 - \partial_x^2 \right]
\phi(x,t) = 0 \quad .
\eeql{eq:wet}

\nid
This paper deals only with the KGE, both because it is
readily related to the Schr\"odinger equation in terms
of which SUSY is usually formulated~\cite{witten,susyqm}, and also because it
(unlike the wave equation) admits NMs which are 
interesting in the present discussion. 

Except for Section~\ref{subsect:tail}, we shall assume that $V(x)$ 
[or $\rho(x) - 1$ in the case of the wave equation (\ref{eq:wet})]
has finite support on the interval $I = [-a,a]$,
which is natural for describing a system of limited extent, surrounded
by a trivial medium such as vacuum.  

We assume that the loss is only due to 
the boundary conditions.  In particular, the potential
$V$ is real.  Absorption may be described by a complex $V$, but causality
then requires dispersion; the necessary generalization~\cite{abs} 
will not be discussed here.

Among the solutions of (\ref{eq:kgt}), we consider 
the space $\Gamma$ of states satisfying the OWC.
We leave the time-domain definition to Section~\ref{sect:orth};
in the frequency domain, a function is in $\Gamma$ if:

\beq
\phi(x) \sim e^{+ i \omega |x|} \quad , \quad |x| \rightarrow \infty
\quad .
\eeql{eq:out1}

\nid
Because $V$ is trivial outside $I$, the asymptotic conditions
(\ref{eq:out1}) can be stated at $x = \pm a$ instead:

\beq
\frac{\phi'(x)}{\phi(x)} = \pm i \omega  \quad , \quad x = \pm a
\quad .
\eeql{eq:out2}

The imposition of two boundary conditions in (\ref{eq:out2})
forces the eigenvalues to be discrete, and these fall
into two classes.  First, there could be bound states
or NMs~\cite{nm2}; these must
(from the Schr\"odinger point of view) have 
$E = \omega^2 < 0$, and hence $\omega$ is purely imaginary.
Since bound-state wavefunctions vanish at infinity, (\ref{eq:out1})
dictates that $\mbox{Im } \omega > 0$.
Second, there could be
QNMs with complex eigenvalues $\omega^2$.
Because these waves escape, $\phi$ decreases, so 
$\mbox{Im } \omega < 0$~\cite{bounds}.  
Provided $\mbox{Re } \omega \ne 0$,
they occur in pairs:
$\omega^{\vphantom{*}}_{-j} = - \omega^*_{j}$, as is readily
shown by conjugating the defining equation and boundary conditions.
Those with $\mbox{Re } \omega = 0$ need not be paired;
these {\em zero modes\/}~\cite{zeromode} will
be of particular importance below. 
Consider the potential shown by the broken line
in Figure~1(a); its NMs and QNMs are shown in Figure~1(b).
In this example, there is one NM (triangle) and a
sequence of QNMs (crosses), including a zero
mode.  In contrast to the Schr\"odinger formulation, 
the use of the KGE and the introduction of $\Gamma$ allows NMs and
QNMs to be discussed together --- and at least in this
example the latter manifestly carry much richer information.

Even though QNM eigenfunctions are not square-integrable
and do not form a conventional Hilbert space,
they are useful for analyzing outgoing waves.
Importantly, the complex QNM frequencies are often directly observable: 
e.g., the central frequency and width of an optical line observed from 
a laser cavity, or the rates of repetition and decay of
a gravitational-wave signal that may within the next 
decade be detected by instruments such as LIGO~\cite{ligo}.
In fact, the spectrum is often so rich that,
under some broad conditions, namely that $V(x)$ vanishes
outside $[-a,a]$ and has singularities at $x=\pm a$, the eigenstates
$\phi_j$ of $\Gamma$ are {\em complete\/} for $x \in I$~\cite{comp}, 
so that for $t\ge0$ the observed wave signal 
can be represented as~\cite{sumnm}

\beq
\phi(x,t) = \sum_{j \in \Gamma} a_j \phi_j(x) e^{-i\omega_j t} \quad .
\eeql{eq:comp}

\nid
In cases where the $\phi_j$'s are not complete, it may still be possible
to characterize the remainder, which could be, for example,
a power law in $t$ for long times~\cite{pricetail,tail}. 
Moreover, when the eigenstates are complete, one can set up a formalism that
closely parallels the conservative case
(see Refs.~\cite{twocomp,bior} and Section~\ref{subsect:qnmnorm}).  
One can even second-quantize using these eigenstates 
as a basis (e.g., to discuss thermal effects and atom--field
interactions in an optical cavity)~\cite{quant}.
These developments have been reviewed~\cite{rmp}.
We shall see that much of the mathematical structure
is preserved under SUSY.




%

\section{Formalism}
\label{sect:form}

\subsection{Supersymmetric quantum mechanics}
\label{subsect:susyqm}

In this paper we consider SUSY
in the one-dimensional KGE (\ref{eq:kgt}),
and especially in the corresponding eigenvalue problem

\beq
H \phi_j(x) = \omega_j^2 \phi_j(x) \quad ,
\eeql{eq:kg}

\nid
where

\beq
H = - \partial_x^2 + V(x) \quad .
\eeql{eq:h1}

\nid
The boundary conditions will be specified later.
In so far as the interest centers on the time-independent
problem (\ref{eq:kg}) and the spectrum, the
Schr\"odinger equation, to which reference is usually 
made~\cite{witten,susyqm}, is
included if we simply relabel $\omega^2 \mapsto E$. 

If there exists another system  

\beq
{\tilde H} = -\partial_x^2 + {\tilde V}(x) \quad ,
\eeql{eq:h2}

\nid
with the same spectrum
(or the same spectrum apart from one state), 
and moreover if the states in the two systems are related by

\beq
{\tilde \phi}(x) = A \phi(x)
\quad ,
\eeql{eq:map}

\nid
where

\bea
A &=& \partial_x + W(x) \quad, \nonumber \\
-A^{\dagger} &=& \partial_x - W(x) \quad ,
\eeal{eq:opa}

\nid
then the
two systems are said to be SUSY partners. 
In particular, if $\phi_j(x)$ is an eigenfunction of $H$,
then ${\tilde \phi}_j(x)$ (provided it does not vanish)
is an eigenfunction of ${\tilde H}$ with the same eigenvalue.
Normalization is deferred to Section~\ref{sect:orth}.

In order for (\ref{eq:map}) to preserve the spectrum, one needs

\beq
AH = {\tilde H}A
\quad ,
\eeql{eq:ahha}

\nid
from which it follows that 

\bea
V(x) &=& W(x)^2 - W'(x) + \Omega^2 \quad , \nonumber \\
{\tilde V}(x) &=& W(x)^2 + W'(x) + \Omega^2 \quad ,
\eeal{eq:vw}

\nid
with $W(x)$ (called the SUSY potential) as in (\ref{eq:opa})
and for some constant $\Omega^2$.  
Since both $V$ and ${\tilde V}$ have to be real, $W$ and
$\Omega^2$ are also real.
Moreover, the Hamiltonians
can be represented as

\bea
H&=& A^{\dagger}A + \Omega^2
\quad , \nonumber \\
{\tilde H} &=& AA^{\dagger} + \Omega^2
\quad .
\eeal{eq:hop}

The two partner systems can be put into one linear space
by introducing Pauli spinors, with $H$ and ${\tilde H}$
associated with $1 \pm \sigma_z$ and $A, A^{\dagger}$
associated with $\sigma_{\mp}$.

Upon reversing the sign of $W$,
(a) $V \leftrightarrow {\tilde V}$ [cf.~(\ref{eq:vw})], and
(b) $A \leftrightarrow -A^{\dagger}$ [cf.~(\ref{eq:opa})];
thus the mapping from
${\tilde H}$ back to $H$ is (up to a sign) achieved by $A^{\dagger}$.
Note however that the mapping is the ``inverse" only in a loose sense:
$A^{\dagger}A$ is not the identity but $H-\Omega^2$
[cf.~(\ref{eq:hop})].

We may regard (\ref{eq:vw}) as a Riccati equation for $W$ in terms
of the given $V$.  For $|x| > a$, both $V$ and ${\tilde V}$
vanish, so $W^2 = -\Omega^2$.  The first-order Riccati equation for $W$
can satisfy two boundary conditions (at $x = \pm a$)
only at special values of $\Omega^2$; this condition becomes 
familiar if we define a generator $\Phi(x)$ by

\beq
W(x) = - \frac{\Phi'(x)}{\Phi(x)} \quad .
\eeql{eq:wphi}

\nid
Then (\ref{eq:vw}) implies

\beq
H \Phi(x) = \Omega^2 \Phi(x) \quad .
\eeql{eq:phieq}

\subsection{Boundary conditions}
\label{subsect:bc}

All the above may be regarded as a review of the familiar
SUSY formalism for the Schr\"odinger equation~\cite{witten,susyqm} if we put
$E = \omega^2$, $E_0 = \Omega^2$ and in particular shift
$V(x)$ so that $E_0 =0$.  Conventionally 
the discussion refers to wavefunctions which vanish
at infinity (or, more precisely, are square-integrable).
Here we consider {\em all\/} eigenfunctions in $\Gamma$,
including both NMs and QNMs,
with the former in the upper and the latter in the lower
half of the frequency plane.

We should check immediately that $\phi \in \Gamma$
implies ${\tilde \phi} \in \Gamma$. 
For $x > a$, if $\phi(x) = C e^{i\omega x}$ then
$
{\tilde \phi} (x) = A \phi(x)
= (i \omega + W_{+} ) C e^{i\omega x}
$,
where 

\beq
W(x{=}{\pm}a) = W_{\pm} \quad 
\eeql{eq:wpm}

\nid  
are the constant values for $x>a$ and $x<-a$ respectively.
Thus $\phi$ and ${\tilde \phi}$ always satisfy the same type
of boundary conditions, and the number of eigenstates in $\Gamma$
is preserved under SUSY: 
$n(\omega)={\tilde n}(\omega)$ [cf.~(\ref{eq:index2a})] --- except
when $A$ or $A^{\dagger}$ destroys a state, to be discussed below.

\subsection{Generator}
\label{subsect:gener}

The various SUSY transformations are related,
in a one-to-one manner, to solutions of (\ref{eq:phieq})
for the generator $\Phi$.
First, suppose $\Omega^2 > 0$, so that $\Omega$ is real.  
Then outside~$I$, $\Phi$ is oscillatory:
either complex (e.g., $e^{i\Omega x}$), 
inadmissible since it leads to a complex $W$; or
real (e.g., $\sin \Omega x $), 
inadmissible since its nodes lead to singularities
in $W$.  Thus, $\Omega^2 <0$, and
we denote $K \equiv |\Omega|$.

At each spatial extreme ($|x| > a $), 
$\Phi$ is in general a sum of increasing and decreasing
functions, i.e.,

\beq
\Phi(x) = c e^{K|x|} + d e^{-K|x|} \quad .
\eeql{eq:mix}

\nid
If both $c, d \neq 0$ (to be called the mixed type), then 
the logarithmic derivative is  
(e.g., for $x > a $)

\beq
W(x) = - K + \frac{2dK}{c} e^{-2Kx} + \cdots \quad , 
\eeql{eq:wasymp}

\nid
so that ${\tilde V} = V + 2W'$ acquires an exponential tail.
(In the special case $\Omega=0$, the tail is
not exponential but asymptotically inverse-square.)  
Thus, if we insist that ${\tilde V}$ 
also has finite support, the mixed type is not allowed
and at each extreme $\Phi$ must be
either purely decreasing 
($\Phi \propto e^{-K|x|}$, denoted as D)
or purely increasing ($\Phi \propto e^{K|x|}$, denoted as I).
Outside $I$, the logarithmic derivative is then
exactly $\pm K$, so $W' = 0$, implying ${\tilde V} = 0$.
When both extremes are considered together,  
$\Phi$ must be one of three types, 
conveniently labelled with the parameter 

\beq
\chi = \sfrac{1}{2} \left[ \mbox{ sign} (W_{+}) 
- \mbox{ sign} (W_{-}) \ \right] \quad ,
\eeql{eq:nu}

\nid
where $\chi = +1$, $-1$, and $0$
respectively for the DD, II, and DI/ID cases (in obvious notation).
In the DI/ID case, the generator is purely incoming from one extreme
and purely outgoing to the other, hence is a total-transmission mode
(TTM).  
The relaxation to allow exponential tails for $V$ and/or 
${\tilde V}$ will be briefly mentioned in Section~\ref{subsect:tail}.

The generator is
annihilated by SUSY: $A \Phi = 0$, 
trivially from (\ref{eq:opa}) and (\ref{eq:wphi}).  
Furthermore, since reversing the sign of $W$
interchanges the partners [cf.\ below (\ref{eq:hop})], in view of
(\ref{eq:wphi}) the transformation from $\tilde{H}$ back to $H$
is generated by $\tilde{\Phi} = \Phi^{-1}$; this is guaranteed
to be an eigenfunction of $\tilde{H}$, also with eigenvalue
$\Omega^2$.
(Despite the notation, ${\tilde \Phi}$ is not
the SUSY partner of $\Phi$: ${\tilde \Phi} \ne A \Phi$.)
The boundary conditions for ${\tilde \Phi} = \Phi^{-1}$ 
interchange D and I, so the reverse transformation
is characterized by ${\tilde \chi} = -\chi$.

These observations allow a simple statement of the
changes in the number of states when 
$H \mapsto \tilde{H}$.  
If $\chi=1$ [cf.\ below (\ref{eq:nu})], an NM $\Phi$ 
is destroyed at $\Omega = iK$ [$\Delta(iK)=1$]
and a QNM $\tilde{\Phi}$ is created at $-\Omega = -iK$
[$\Delta(-iK) = -1$].
If $\chi=-1$, a QNM $\Phi$ 
is destroyed at $\Omega = -iK$ [$\Delta(-iK)=1$]
and an NM $\tilde{\Phi}$ is created at $-\Omega = iK$
[$\Delta(iK) = -1$].
If $\chi =0$, no eigenstates of $\Gamma$ are created or
destroyed [$\Delta(iK)=\Delta(-iK)=0$], since $\Phi$ and
$\tilde{\Phi}$ are TTMs rather than NMs or QNMs.  Thus, all three cases
satisfy

\beq
\Delta(iK) = -\Delta(-iK) = \chi \quad ,
\eeql{eq:index3}

\nid
where we emphasize the convention $\Omega = \pm iK$ with $K>0$.
The cases $\chi = \pm 1$ lead to 
Hamiltonians whose spectra in $\Gamma$ differ by one state
(said to be essentially isospectral), whereas the case $\chi = 0$ 
leads to Hamiltonians whose spectra in $\Gamma$ are identical
(said to be strictly isospectral).

These remarks provide a more complete picture of the mapping 
of eigenstates in $\Gamma$ under SUSY: states
do not simply appear/disappear, but are mapped to the mirror
point in the complex plane.

Not all NMs, QNMs or TTMs
are eligible as the generator.  First, $\Omega^2$ must be real,
which restricts QNMs to zero modes.  Secondly, $\Phi$ cannot
have nodes, or else $W$ would acquire singularities.  In the
case of NMs, this restricts $\Phi$ to the ground state.  For QNMs,
nodes are not required by general theorems.  At least
for repulsive potentials,   
each eigenfunction can have {\em at most\/}
one node or antinode; thus, 
for a symmetric repulsive $V$, even-parity eigenfunctions
can have no nodes.  There is consequently much more freedom in choosing
a QNM (as opposed to an NM) as the generator.
Some general statements concerning nodes in QNMs are
given in Appendix~\ref{sect:appnode}.  



%

\section{Orthonormality}
\label{sect:orth}


\subsection{Orthonormality for NMs}
\label{subsect:orthnm}

For conservative systems, SUSY preserves orthonormality.
There are two issues:
orthogonality is preserved because the transformed
NMs are eigenvectors of the self-adjoint operator 
${\tilde H}$; and  
normalization is preserved if the transformation
is changed to

\beq
\phi_j \mapsto {\underline {\tilde \phi}}{}_j
= N_j {\tilde \phi}_j  = N_j A \phi_j
\quad ,
\eeql{eq:ortho01}

\nid
with 

\beq
N_j^{-2} = \frac{
\langle {\tilde \phi}_j | {\tilde \phi}_j \rangle }
{ \langle \phi_j | \phi_j \rangle }
\quad .
\eeql{eq:ortho02}

\nid
Eq.~(\ref{eq:ortho01}) applies to each eigenstate $j$ 
other than the generator $\Phi$ itself,
namely the ground state.
It is readily shown that $N_j^{-2} = \omega_j^2 - \Omega^2$,
a result that can also be read off as a special case
of the derivation below for states in $\Gamma$.  Since
$\omega_j^2 - \Omega^2$ is the eigenvalue of
$A^{\dagger} A$ [see (\ref{eq:hop})], 
(\ref{eq:ortho01}) can be written in the operator form

\beq
\phi \mapsto {\underline {\tilde \phi}}
= A \left( A^{\dagger} A \right)^{-1/2} \phi
\quad ,
\eeql{eq:ortho03}

\nid
valid for any state $\phi$, not just
frequency eigenfunctions.
(This and similar formulas below are restricted to
the subspace orthogonal to $\Phi$.)
This makes it formally easy to verify the preservation of inner products:

\bea
\langle {\underline {\tilde \psi}} | {\underline {\tilde \phi}} \rangle
&=&
\langle \psi | (A^{\dagger}A)^{-1/2} \, A^{\dagger} A \, 
\, (A^{\dagger}A)^{-1/2} |  \phi \rangle
\nonumber \\
&=& \langle \psi | \phi \rangle 
\quad .
\eeal{eq:ortho04}

\nid
However, when operating on a
general wavefunction $\phi$, the factor $(A^{\dagger}A)^{-1/2}$
can only be evaluated by projecting $\phi$ onto the eigenfunctions,
and scaling each component by $N_j$.  Thus, in practice,
the significant result is the evaluation
of this factor.
We now generalize these concepts to states in $\Gamma$, 
in particular QNMs.


\subsection{Normalization and inner product for QNMs}
\label{subsect:qnmnorm}

It is necessary to digress and review the
concepts of orthogonality and normalization for QNMs.
The central issue is that with the OWC, 
$H$ is {\em not\/} self-adjoint
in the usual sense, and different QNMs are
{\em not\/} orthogonal under the usual inner product.  Likewise,
the norm $\int |\phi_j|^2 dx$ is divergent,
since the wavefunction grows exponentially
at infinity.

An appropriate normalizing factor for QNMs was first introduced
by Zeldovich~\cite{zel}, and later generalized 
and applied to other situations~\cite{pert1}, including models of
linearized waves propagating on a Schwarzschild background~\cite{dirt}:

\bea
( \phi_j , \phi_j ) &=&
 2 \omega_j \int_{-a}^{a} \phi_j(x)^2 \, dx
\nonumber \\
&&{} + i \left[ \phi_j(-a)^2 + \phi_j(a)^2 \right]
\quad .
\eeal{eq:ortho05}

\nid
This expression goes as $\phi_j^2$ rather than
$|\phi_j|^2$, and is in general not real.
The limits of the integral and the surface terms
can be shifted from $\pm a$ to any $b_{\pm}$,
where $\pm b_{\pm} > a$, without affecting the
value of (\ref{eq:ortho05}).    
This definition also applies to NMs:
the surface terms vanish if we 
take $b_{\pm} \rightarrow \pm \infty$, 
recovering the conventional norm apart from a factor of $2\omega_j$.
In the QNM case, (\ref{eq:ortho05}) is the correct
normalizing factor in the sense that, e.g., under a
perturbation $V \mapsto V + \Delta V$, the complex QNM eigenvalues
change by

\beq
\Delta (\omega_j^2) = 
\frac{ \int \phi_j(x)^2 \Delta V(x) dx }
{ (\phi_j , \phi_j ) }
\quad .
\eeql{eq:ortho06}

\nid
Since one no longer has positivity, there is the possibility
that $(\phi_j , \phi_j) =0$.
This exceptional case can be separately
taken care of~\cite{jordan}, and some interesting
aspects are dealt with in Section~\ref{sect:jb}. 

To go beyond the normalizing factor and
discuss an analog of orthogonality, one has to first
regard each state as a two-component vector 
$\bbox{\psi} = ( \psi^1 , \psi^2 )^{\rm T} \equiv 
( \psi, \partial_t \psi )^{\rm T}$,
which is most easily motivated by noticing that the
dynamics requires two sets of initial data.
The space $\Gamma$ is then defined as all $\bbox{\psi}$ satisfying

\beq
\psi^2(\pm a) = \mp \partial_x \psi^1(\pm a) \quad.
\eeql{eq:defgam}

\nid
For an eigenstate, 
$\bbox{\phi}_j = (\phi_j, -i\omega_j \phi_j)^{\rm T}$,
and (\ref{eq:out2}) would follow from (\ref{eq:defgam}).

In terms of the two-component vector, 
one can define a bilinear map~\cite{twocomp,rmp}

\bea
( \bbox{\psi} , \bbox{\phi} )
&=& i \left\{ \int_{-a}^{a} \left[ \psi^1(x) \phi^2(x) +
\psi^2(x) \phi^1(x) \right] dx \right.
\nonumber \\
&&{}+\left. \vphantom{\int_{-a}^{a}}
\left[ \psi^1(-a) \phi^1(-a) + \psi^1(a) \phi^1(a) \right] \right\}
\quad ,
\eeal{eq:ortho13}

\nid
to take the place of the usual
inner product.  For an eigenfunction,
$( \bbox{\phi}_j , \bbox{\phi}_j )$ agrees with (\ref{eq:ortho05}).
The dynamics can be written in the first-order form
$i\partial_t \bbox{\phi} = {\cal H} \bbox{\phi}$,
with 

\beq
{\cal H} = i \pmatrix{ 0 & 1 \cr \partial_x^2 - V & 0 }
\quad .
\eeql{eq:ortho14}

\nid

Importantly, ${\cal H}$ is symmetric:

\beq
( {\cal H} \bbox{\psi} , \bbox{\phi} )
= ( \bbox{\psi} , {\cal H} \bbox{\phi} )
\quad ,
\eeql{eq:ortho07}

\nid
in the proof of which the surface terms generated
upon integration by parts exactly cancel against
those in (\ref{eq:ortho13}).
The relation
(\ref{eq:ortho07}) is the analog of self-adjointness, and
leads to the usual proof that
for two eigenvectors,

\beq
( \bbox{\phi}_k , \bbox{\phi}_j ) = 0
\eeql{eq:ortho08}

\nid
whenever $\omega_k \neq \omega_j$.  Provided that $( \bbox{\phi}_j , \bbox{\phi}_j ) \neq 0$~\cite{jordan},
one can normalize these eigenfunctions in the usual way,
i.e., by requiring (\ref{eq:ortho08}) to be 
$2 \omega_k \delta_{kj}$ in general 
[cf.~(\ref{eq:ortho05}) for this factor].
We henceforth refer to this
property as orthonormality [and to (\ref{eq:ortho08}) alone
as orthogonality].  It also follows trivially that, provided
this orthonormal system is complete
(which is the case under fairly broad assumptions; see
Section~\ref{subsect:qnm}), time evolution is given by

\beq
\bbox{\phi}(t) = \sum_j a_j \bbox{\phi}_j \, e^{-i\omega_j t}
\quad ,
\eeql{eq:ortho09}

\nid
generalizing (\ref{eq:comp}) to two components,
and

\beq
a_j = \frac{ (  \bbox{\phi}_j , \bbox{\phi}(t{=}0) ) }
{ ( \bbox{\phi}_j , \bbox{\phi}_j ) }
\quad .
\eeql{eq:ortho10}

\nid
The preservation of 
orthonormality under SUSY should therefore be sought in terms of
the bilinear map (\ref{eq:ortho13}).


\subsection{Normalized SUSY transformation for QNMs}
\label{subsect:orthqnm}

We first present a derivation
of orthonormality that does not explicitly
require the two-component formalism.
With orthogonality already guaranteed by (\ref{eq:ortho08}),
it remains to compute the normalizing factor

\bea
({\tilde \phi}_j , {\tilde \phi}_j )
&=& 2\omega_j  \int_{-a}^{a} 
\left[ (\partial_x + W) \phi_j \right]^2 dx
\nonumber \\
&&{} + i \left[ {\tilde \phi}_j(-a)^2 + {\tilde \phi}_j(a)^2 \right]
\quad.
\eeal{eq:ortho11}

\nid
Integrate by parts to convert 
$(\partial_x \phi)^2$ to $-(\partial_x^2 \phi)\phi$
plus a surface term,
express the second derivative in terms of $V - \omega_j^2$
by means of the eigenvalue equation, and write the potential
as $V = W^2 - W'+ \Omega^2$.  Then, apart from 
a term $\propto \omega_j^2 - \Omega^2$,
the integrand becomes a total derivative $\partial_x(W\phi^2)$.  
Using $W(\pm a)^2 = -\Omega^2$
and $\partial_x \phi_j (\pm a) = \pm i \omega_j \phi_j(\pm a)$
then leads to

\beq
\frac{ ({\tilde \phi}_j , {\tilde \phi}_j ) }
{(\phi_j , \phi_j ) }
= \omega_j^2 - \Omega^2   
\quad .
\eeql{eq:ortho12}

\nid
Incidentally, the conservative case (nodal conditions 
at the ends of the interval~\cite{nm2})
is recovered by simply dropping all surface terms.

Since the ratio (\ref{eq:ortho12}) is the eigenvalue of 
$A^{\dagger}A$, we can
again write the normalized transformation 
for each eigenfunction as (\ref{eq:ortho03}).


\subsection{SUSY for two-component form}
\label{subsect:twocomp}

For eigenstates, the two components are trivially related, but
in order to perform SUSY transformations on general
wavefunctions in $\Gamma$ (e.g., given a time-dependent state,
to find its partner at all times), the second
component must be considered explicitly.

Since SUSY must commute with time evolution
and $\phi^2 = \partial_t \phi^1$, both
components must transform in the same way.
Thus, the (unnormalized) SUSY transformation on two-component vectors is
${\cal A} = \mbox{diag } ( A , A )$, which satisfies

\beq
( {\cal A} \bbox{\psi} , \bbox{\phi} )
= ( \bbox{\psi} , {\cal A}^{\dagger} \bbox{\phi} )
\quad ,
\eeql{eq:ortho17}

\nid
where ${\cal A}^{\dagger} \equiv \mbox{diag } ( A^{\dagger} , A^{\dagger} )$.
In deriving (\ref{eq:ortho17}), one has to integrate 
by parts: $\partial_x$ changes sign so that 
${\cal A}$ turns into ${\cal A}^{\dagger}$.
The surface terms are seen to work out by using,
e.g., $\phi^2(a) = \partial_t \phi^1(a)
= -\partial_x \phi^1(a)$, and the known values of $W_{\pm}$.
Note that ${\cal A}^{\dagger} {\cal A} = ( H - \Omega^2 ) \openone$,
${\cal A} {\cal A}^{\dagger}  = ( {\tilde H} - \Omega^2 ) \openone$,
i.e., the products do not relate to the two-component
${\cal H}$.

With (\ref{eq:ortho17}), it is 
straightforward to show that the normalized SUSY transformation

\beq
\bbox{\phi} \mapsto \underline{\tilde {\bbox{\phi}}} =
{\cal A} \, ( {\cal A}^{\dagger} {\cal A} )^{-1/2} \bbox{\phi}
\quad ,
\eeql{eq:ortho18}

\nid
defined on the subspace orthogonal to ${\bbox {\Phi}}$,
preserves the bilinear map, in a manner that exactly parallels
(\ref{eq:ortho04}).  In the exceptional case of SUSYs that generate
a doubled state (see Section~\ref{sect:jb}), the subspace
has to exclude the {\em two\/} states on which
$H - \Omega^2$ vanishes.

The linear-space structure for open systems
(e.g., the replacement of inner products by bilinear
maps) has an intrinsic geometric meaning for all outgoing states,
not just QNMs~\cite{factor}.  It is therefore pleasing
that this structure is preserved by SUSY, 
a superficially unrelated concept.



%

\section{Jordan Blocks}
\label{sect:jb}

A key concept in SUSY is the preservation of the
spectrum.  However, dissipative systems (such as
waves satisfying the OWC) admit a spectral
property not found for conservative systems.  In terms
of the Wronskian $J(\omega)$ to be defined below,
this is exhibited as an $M$th-order zero ($M>1)$.
Such a multiple zero emerges naturally when
$M$ QNM eigenvalues coalesce
as system parameters are tuned, so that
$M-1$ eigenvectors are ``lost"~\cite{oneqnm} and must be replaced by other 
degrees of freedom.  
Thus, the Hamiltonian cannot be written as a diagonal
matrix in the (biorthogonal) basis of eigenstates,
but can only be decomposed into (Jordan) blocks~\cite{jb0}
of size $M \times M$.  When this happens,
$( \bbox{\phi}_j , \bbox{\phi}_j )$ will vanish for some $j$, invalidating  
the formalism in Section~\ref{sect:orth} 
[see, e.g., (\ref{eq:ortho06}) and (\ref{eq:ortho10})].
These issues have been discussed in detail with reference
to waves in open systems~\cite{jordan}.  

The simplest example of a JB (with $M=2$) is
a harmonic oscillator going through critical damping.
The eigenvalues $\omega_{\pm} = \pm \omega_{\rm R} -i\gamma$
coalesce when $\omega_{\rm R} \rightarrow 0$.  With one
eigenvalue lost, the dynamics is not given by a sum of 
eigenfunctions with time dependence $\exp(-i\omega_{\pm} t)$, but by only
{\em one\/} such eigenfunction, 
plus another term whose time dependence carries a prefactor $t$.

Our purpose in this Section
is to establish that SUSY maps a JB in $H$ into a JB in ${\tilde H}$,
preserving the order $M$ except when the eigenvalue coincides
with $\pm \Omega$.  


\subsection{Wronskian}
\label{sect:wron}

In this subsection we introduce the Wronskian $J(\omega)$,
define JBs in terms of its multiple
zeros and describe the mapping of JBs under
SUSY by a relation between $J(\omega)$ and its
counterpart ${\tilde J}(\omega)$.

In the
original system $H$, define solutions of the wave equation $f(\omega,x)$
and $g(\omega,x)$ satisfying the boundary conditions
$
f(\omega,-a) = 1, \;
f'(\omega,-a) = -i\omega, \; 
g(\omega,a) = 1, \;
g'(\omega,a) = i\omega,$
where $\, {}' = \partial_x
$.  The function values are arbitrary normalizations, 
while the derivatives impose the OWC
on the left and right respectively.
An eigenstate $\phi_j$ in $\Gamma$  
satisfies the boundary condition on {\em both} the left
(as for $f$) and the right (as for $g$):
$\phi_j \propto f(\omega_j,x) \propto g(\omega_j,x)$.
Thus, the zeros of the (position-independent) Wronskian 

\beq
J(\omega) = f'(\omega,x) g(\omega,x) - f(\omega,x) g'(\omega,x)
\eeql{eq:doub04}

\nid
identify the eigenvalues in $\Gamma$.  It can be shown [see
(\ref{eq:fgJ}) below] that 
$(\phi_j , \phi_j) \propto dJ(\omega_j)/d\omega$, so an 
$M$th-order zero of $J$ ($M>1$) corresponds to the
generalized norm being zero, and is precisely the JB phenomenon
that we wish to investigate.

It is natural to generalize the definition of $n(\omega)$
to be the order of the zero, {\em viz.\/}

\beq
n(\omega) = \frac{1}{2\pi i} \oint 
\frac{dJ(\omega')/d\omega'}{J(\omega')} d\omega'
\quad ,
\eeql{eq:n-cont}

\nid
on a contour of winding number $+1$ enclosing $\omega$.
This definition makes it clear that the total number of states 
(but {\em not\/} necessarily of eigenstates)
within a contour is preserved under continuous changes of the system
parameters.  We note for future reference (see Section~\ref{subsect:tail})
that poles of $J$ (which can only occur if $V$ does not
have finite support) count as negative values of $n$.

Now consider the analogous construction in the partner
system ${\tilde H}$, obtained for example by using an NM $\Phi$
of $H$ as the generator, i.e., for $\chi = 1$.  By our convention,
$\Phi$ is associated with a frequency $\Omega = iK$,
and $W_{\pm} = \mp i \Omega$.  The SUSY transformation gives~\cite{fgnote}

\bea
{\tilde f}(\omega,x) &=& (\partial_x + W) f(\omega,x)
\nonumber \\
{\tilde g}(\omega,x) &=& (\partial_x + W) g(\omega,x)
\quad ,
\eeal{eq:doub05}

\nid
leading to the Wronskian

\beq
{\tilde J}^{\rm u}(\omega) 
= 
{\tilde f}'(\omega,x){\tilde g}(\omega,x) - 
{\tilde f}(\omega,x){\tilde g}'(\omega,x)
\quad .
\eeql{eq:doub07}

When (\ref{eq:doub07}) is written out using (\ref{eq:doub05}), 
some terms cancel by using
$J' = 0$, the second derivatives can be eliminated
by the defining equation, and $V$ is expressed in terms
of $W$ and $\Omega^2$.  Some arithmetic then leads to

\beq
{\tilde J}^{\rm u}(\omega) =
(\omega^2-\Omega^2)\,J(\omega)
\quad.
\eeql{eq:doub08a}

\nid
This Wronskian is however unnormalized (as indicated by the
superscript), since only $C{\tilde f}$ and $D{\tilde g}$ satisfy the
normalization conventions at $-a$ and $+a$ respectively,
where 
$ C = -D = i(\omega - \Omega)^{-1}$.  Thus the normalized
Wronskian ${\tilde J}(\omega) =
CD {\tilde J}^{\rm u}(\omega)$ is

\beq
{\tilde J}(\omega)
= \frac{\omega + \Omega} {\omega - \Omega} \, J(\omega)
\quad .
\eeql{eq:doub08}

The central result (\ref{eq:doub08}) neatly summarizes the correspondence
between the two spectra for $\chi = 1$.  [Similar formulas
for the other cases can all be consolidated by changing
$\Omega \mapsto i\chi K$ in (\ref{eq:doub08}), and 
will not be separately discussed.]
(a) For $\omega \ne \pm \Omega$,
the spectra of $H$ and ${\tilde H}$ are the same:
a JB of order $M$ in $H$ maps to
a JB also of order $M$ in ${\tilde H}$
at the same frequency.
This should be no surprise since up to the point of coalescence
(for $M \ge 2$),
the eigenvalues of the two systems are guaranteed  
(cf.\ Section~\ref{sect:form}) to be in one-to-one
correspondence; in a sense, the result here merely demonstrates
that the limit is not singular.  
(b) Moreover, {\em at\/} the special frequencies $\pm \Omega$,
$M$ states at $\Omega$ (of which only
{\em one\/} is an eigenstate) are mapped into $M{-}1$ states 
at $\Omega$ plus one state at $-\Omega$.
In other words, we recover (\ref{eq:index3})  
even for JBs, i.e., even when $n, {\tilde n} > 1$.
Anticipating the possibility of poles in $J$
(cf.~Section~\ref{subsect:tail}), we note that (\ref{eq:doub08})
implies that such poles
are also preserved under SUSY, and 
would be accommodated by 
(\ref{eq:index2a}), (\ref{eq:index2}), and (\ref{eq:index3}) with 
negative $n$, ${\tilde n}$.

Incidentally, (\ref{eq:ortho12}) on the change
in normalization under SUSY follows 
simply from (\ref{eq:doub08a}), 
since the bilinear map is related to the Wronskian by~\cite{rmp,normgen}

\beq
(f(\omega_j),g(\omega_j))=
-\left[ \frac{d J (\omega)} {d \omega} \right]_{\omega_j}
\quad .
\eeql{eq:fgJ}


\subsection{Doubling of states by SUSY}
\label{sect:double}

We have seen that (say for $\chi = 1$) one has 
${\tilde n}(-iK) - n(-iK) = 1$.
Where this increases from 0 to 1, the situation
is straightforward --- a QNM is created.  When the
increase is from 1 to 2, the situation is more subtle
and merits a detailed examination. 
[The general case where this increases from $M$ to $M{+}1$
($M \ge 1$) will not be shown.]

Consider a system $H$ with an NM $\Phi$ at 
$\Omega = iK$ and accidentally also a QNM $\Psi_j$ at $-\Omega'
\approx -\Omega$.
If $\Omega' \ne \Omega$, there are {\em two\/}
corresponding QNMs in the ${\tilde H}$-system, namely
${\tilde \Phi} = \Phi^{-1}$ at $-\Omega$
and ${\tilde \Psi_j} = A \Psi_j$ at $-\Omega'$.
Now tune the parameters of $H$ so that $\Omega' \rightarrow \Omega$;
in the limit we must have 
${\tilde \Psi_j} \propto {\tilde \Phi}$\cite{jordan,oneqnm},
as is readily verified.
The proportionality constant can be evaluated by

\bea
\frac{\tilde{\Psi}_j}{\tilde{\Phi}}
&=& \frac{\tilde{\Psi}_j(-a)} {\tilde{\Phi}(-a)}
=2i \Omega\Psi_j(-a) \Phi(-a)
\nonumber\\
&=& \frac{\tilde{\Psi}_j(a)} {\tilde{\Phi}(a)}
= -2i\Omega\Psi_j(a)\Phi(a)
\quad.
\eeal{eq:doub01}

\nid
The agreement of these two expressions can also be 
seen without invoking SUSY.  One notes that $\bbox{\Phi}$ 
and $\bbox{\Psi}_j$, being eigenfunctions of ${\cal H}$
with distinct eigenvalues, are orthogonal.  In the 
bilinear map, the integral term vanishes because
the frequencies are opposite,
leaving only the surface terms.  Thus one finds
$0=(\vec{\Psi}_j,\vec{\Phi})=i[\Psi_j(-a)\Phi(-a)+\Psi_j(a)\Phi(a)]$.

Now the frequency $-\Omega$ in ${\tilde H}$ must be associated 
with a doubled state.  This can be seen in two ways: (a)
until the limit $\Omega' = \Omega$, there are two distinct states;
(b) from the key relation (\ref{eq:doub08}), ${\tilde J}$
has a second-order zero.  With the two QNMs collapsed
into one, there has to be another basis vector, to which we now turn.

Using the normalization of $f$, one has 
$\Psi_j(x)=\Psi_j(-a)f(-\Omega,x)$, implying 
$\tilde{\Psi}_j(x)=\Psi_j(-a) \tilde{f}(-\Omega,x)$.
[Analogous formulas for $\Phi$ and $\tilde{\Phi}$ 
follow from (\ref{eq:doub01}).] 
The double zero of $\tilde{J}(\omega)$ at $\omega=-\Omega$ 
means that $\tilde{f}(\omega,x)$ satisfies the OWC at $x=a$ 
not only {\em at\/} $\omega = -\Omega$, but also to first order away from
the zero.  This makes it plausible that, in the QNM expansion, 
$\partial_\omega\tilde{f}(\omega,x)|_{-\Omega}$ 
takes the place of the ``missing" eigenfunction when 
$\tilde{\Phi}$ and $\tilde{\Psi}_j$ coincide, 
which has been confirmed in detail\cite{jordan}.  One thus defines,
for an arbitrary $\alpha$, a pair of functions ${\tilde \Psi}_{j,n}$,
where $n=0,1$ is an intra-block index:

\bea
\tilde{\Psi}_{j,0}(x) &=& \tilde{\Psi}_j(x)
\nonumber \\
\tilde{\Psi}_{j,1}(x) &=&
\Psi_j(-a) \partial_\omega \tilde{f}(\omega,x) |_{-\Omega}
+ \alpha \tilde{\Psi}_j(x)
\quad,
\eeal{eq:Psi01}

\nid
and the second function satisfies

\beq
( {\tilde H}- \Omega^2 ) 
{\tilde \Psi}_{j,1} = -2 \Omega {\tilde \Psi}_{j,0}
\quad.
\eeql{eq:WEJB}

\nid
Using this, one verifies that an outgoing 
solution is given by 
${\tilde \Psi}_{j,1}(t) \equiv
( {\tilde \Psi}_{j,1}-it {\tilde \Psi}_{j,0} )
e^{i\Omega t}$.  This time dependence shows that the 
associated second component should be
${\tilde \Psi}_{j,1}^2 = 
-i ( -\Omega {\tilde \Psi}_{j,1}
+ {\tilde \Psi}_{j,0} )$. 
The prefactor $t$ and its counterpart if any in the $H$-system
will be further explored below.

Next consider the bilinear map and normalization for these functions.
For a double zero one always has 
$( \tilde{\vec{\Psi}}_{j,0}, \tilde{\vec{\Psi}}_{j,0})=0$; 
cf.~(\ref{eq:fgJ}) 
[a choice of $\alpha$ in (\ref{eq:Psi01}) such that also
$( \tilde{\vec{\Psi}}_{j,1}, \tilde{\vec{\Psi}}_{j,1})=0$
would be useful in wavefunction expansions].
The JB is normalized by one overall factor, 
the bilinear map between the two basis states~\cite{jordan}:

\bea
\frac{ ( \tilde{\vec{\Psi}}_{j,1}, \tilde{\vec{\Psi}}_{j,0}) }
{(\vec{\Psi}_j,\vec{\Psi}_j)}
&=&
\left[ \frac{-\sfrac{1}{2} d^{2}\tilde{J}^{\rm u}(\omega)/d\omega^{2} }
{-dJ(\omega)/d\omega} \right]_{-\Omega}
\nonumber \\
&=& -2\Omega
\quad ,
\eeal{eq:normjb}

\nid
where in the numerator we have used the result analogous to 
(\ref{eq:fgJ}) for a double zero. 

Finally the reverse transform generated by 
$-A^{\dagger}$~\cite{missing} satisfies 
the following properties.
(a) $A^{\dagger}{\tilde \Psi}_{j,0} \propto A^{\dagger}{\tilde \Phi}=0$. 
(b) Hence in $A^{\dagger} {\tilde \Psi}_{j,1}(t)$, 
the term $\propto te^{i\Omega t}$ is annihilated.
(c) The remaining term in $A^{\dagger} {\tilde \Psi}_{j,1}$
is $c \Psi_j$,
readily seen by observing that

\bea
(H-\Omega^2) (A^{\dagger}{\tilde \Psi}_{j,1})
&=&
A^{\dagger}({\tilde H}-\Omega^2) {\tilde \Psi}_{j,1}
= A^{\dagger}(-2\Omega {\tilde \Psi}_{j,0} ) 
\nonumber \\
&=& 0
\quad ,
\eeal{eq:jb02}

\nid 
so that $A^{\dagger}{\tilde \Psi}_{j,1}$ is an eigenfunction
of $H$ with eigenvalue~$\Omega^2$.  In particular, the time dependence
is $e^{i\Omega t}$ without any prefactor $t$.
A straightforward computation shows that $c = -2\Omega$.

Since $A^{\dagger}A$ is not the identity,
two different states in the $H$-system
can be associated with ${\tilde \Psi}_{j,1}$.  The first
is $A^{\dagger}{\tilde \Psi}_{j,1}$ as discussed above.
The second is the SUSY pre-image of ${\tilde \Psi}_{j,1}$ under $A$,
which is readily found by noticing that~\cite{notunique}

\bea
{\tilde \Psi}_{j,1}(x) &=& 
\Psi_j(-a) \times \partial_\omega [(\partial_x+W)f(\omega,x) ]_{-\Omega}
\nonumber \\
&=& A \, \Psi_j(-a)\partial_\omega f(\omega,x)|_{-\Omega}
\quad.
\eeal{eq:doub21}

\nid
However, $\Psi_j(-a)\partial_\omega f(\omega,x)|_{-\Omega}$
is {\em not\/} outgoing, since $J(\omega)$
only has a first-order zero at $-\Omega$, 
and consequently $f(\omega,x)$ satisfies the OWC
at $x=+a$ only {\em at\/} $-\Omega$, but not to first
order away from it.  

These remarks
completely resolve the puzzle related to the prefactor $t$
in the time evolution in the ${\tilde H}$-system.  Namely,
of the two corresponding wavefunctions of $H$, 
one has an exponential time dependence, while the other is not in $\Gamma$.



%
\section{Examples}
\label{sect:ex}

The formalism developed in this paper is general, in
that given $V(x)$ with finite support, SUSY
partners can be constructed whenever there are 
nodeless generator candidates.
Nevertheless, some simple examples will
suffice for an illustration.

Let $V$ be a square barrier
of height $V_0$ on $I$; without loss of generality henceforth $a=1$.  
In the even sector there 
are two zero modes for small values of $V_0$; e.g., for
$V_0 = 0.16$ they occur at 
$\omega_1 = -0.181i$, $\omega_2 = -2.500i$.
The wavefunctions are $\phi_j(x) = \cosh \alpha_j x$ within $I$
(with $\alpha_1 = 0.242$, $\alpha_2 = 2.506$),
and a real exponential for $|x| > a$; clearly each
$\phi_j$ has no nodes. 

This example already illustrates, as unexceptional,
the existence of several nodeless zero modes ---
any one of which can be used as the generator $\Phi$.
We choose the state at $\Omega = \omega_1$;
since $\Phi$ is purely increasing at both
extremes, $\chi = -1$.
Figure~1(a) shows $V$ (solid line)
and ${\tilde V}$ (dotted line).  Figure~1(b) shows the spectra
in the complex $\omega$-plane:
most eigenvalues are common (crosses); 
one QNM exists only in $H$ (circle), while 
one NM exists only in ${\tilde H}$ (triangle) ---
the two systems are essentially isospectral.
Another essentially isospectral partner can be constructed using the state
at $\omega_2$ as the generator.  The reverse transformations
are characterized by ${\tilde \chi} = 1$.

As $V_0$ increases, the two zero modes move closer together and
merge at $V_0 = 0.291$,
forming a JB in $H$ (with $M=2$).  This example  
illustrates the general results of Section~\ref{sect:jb}
(with $H$ and ${\tilde H}$ interchanged).

As another example, consider a symmetric multi-step square barrier, with
$V(x)$ being $-10$ for $|x| < 0.1$, $1$ for $0.1 < |x| < 1$ and
zero for $|x| > 1$ [solid line in Figure~2(a)].  In this example, there
is a TTM at $\Omega = -0.990i$ [square in Figure~2(b)] --- in fact
a parity doublet with one propagating from the left (TTM${}_{\rm L}$)
and one propagating from the right (TTM${}_{\rm R}$).
We choose the former as the generator for a $\chi = 0$ transformation.
The partner potential ${\tilde V}$ is shown by the broken line in Figure~2(a).
Since the generator is not symmetric, neither is ${\tilde V}$.
The states in $\Gamma$ are exactly preserved [crosses in Figure~2(b)].

In this example it is interesting to consider not just the states
in $\Gamma$, but also TTMs (see also Section~\ref{subsect:ttm}).  
By arguments similar to those in
Section~\ref{sect:form}, one TTM${}_{\rm L}$ $\Phi$ is destroyed
and one TTM${}_{\rm R}$ ${\tilde \Phi} = \Phi^{-1}$
is created at the same frequency $\Omega$
[square in Figure~2(b)].  However, in this example, because $V$
is symmetric, there is also a TTM${}_{\rm R}$ $\Psi(x) = \Phi(-x)$
at $\Omega$, and its partner ${\tilde \Psi}
= A \Psi$ is again a TTM${}_{\rm R}$ in the ${\tilde H}$-system.  
Thus, in the ${\tilde H}$-system, there is a {\em doubled\/}
TTM${}_{\rm R}$ state at $\Omega$.  The situation can again
be analysed in terms of the double zero of a Wronskian 
${\tilde J}_{\rm R}(\omega)$,
but in this case $J_{\rm R}$ and ${\tilde J}_{\rm R}$ refer to TTM${}_{\rm R}$'s.
For example, $J_{\rm R}$ is now defined in terms of a function
$f$ satisfying the outgoing-wave condition at $x=-a$ and
a function $g$ satisfying the incoming-wave condition
at $x=a$.  The obvious adaptation of the discussion
in Section~\ref{sect:jb} will not be given.



%
\section{Discussion}
\label{sect:disc}

\subsection{Summary}
\label{subsect:summ}

In this paper, we have extended the usual discussion
of SUSY as a relation between NMs of partner systems
to include the QNMs as well. By viewing all these together
in the space $\Gamma$, a more complete picture emerges.
For example, in the usual discussion for NMs only, 
essentially isospectral transformations are said to
lose or gain one state; 
now we see that (for $\chi = \pm 1$)
when an NM appears (disappears), a corresponding QNM
disappears (appears).

Furthermore, we have shown that the nontrivial
linear-space structure for QNMs and any possible 
JBs are preserved by SUSY --- the latter being a feature
not found in conservative systems.

QNMs differ from NMs in two further regards.  First, they have
complex frequencies; nevertheless, even with 
twice as many constraints, matching the spectra
turns out to be not any more difficult.  Second, QNMs need not have
an increasing number of nodes, and it is often possible
to find several nodeless QNMs which generate
distinct SUSY transformations --- whereas the analogous
operation for NMs would restrict the generator to the nodeless ground state.

These wider perspectives are
gained only because attention is paid to the Klein--Gordon
rather than the Schr\"odinger equation, since the
concept of outgoing waves has no meaning in an equation
that is first-order in time.  

Two further important properties are also preserved. 
(a)~If $V$ has a singularity say at 
$x=\pm a$ (e.g., a step), 
then ${\tilde V}$ will have the same type of singularity, 
but with opposite sign, as can be seen from (\ref{eq:vw}) 
by noticing that the most singular part is $W'$. 
(b)~If $V$ has finite support, 
then provided $\Phi$ is not of the mixed type, ${\tilde V}$ 
would likewise have no tail. These two properties are precisely the conditions 
for the eigenstates in $\Gamma$ to be complete~\cite{rmp}.  
Thus, SUSY maps a complete basis to a complete basis 
(if, for the $\chi = \pm 1$ cases, $\Phi \mapsto {\tilde \Phi}$ 
is included as well).

\subsection{Inversion}
\label{subsect:inv}

This work partially answers the
question of QNM {\em inversion\/}.  It is well known~\cite{invert}
that on a finite interval, {\em two\/} sets of real NM frequencies
uniquely determine the potential $V$.  Does 
{\em one\/} set of {\em complex\/} eigenfrequencies in $\Gamma$ 
also uniquely determine $V$?  The answer is negative:
there can be strictly isospectral potentials if 
a TTM with purely imaginary frequency exists
and can be used as a $\chi = 0$ generator; Figure~2
provides one such example.  
However, the following scenario is not yet ruled
out.  If we consider a half-line problem $0 < x < \infty$
(say corresponding to the radial variable in a 3-d system),
imposing a nodal condition at $x=0$ and the OWC for $x > a$, can 
one set of QNM frequencies uniquely determine the potential?
SUSY transformations (\ref{eq:map}) do not directly resolve
this possibility --- for which there is some numerical 
evidence~\cite{numinv2} --- since these one-sided systems
do not feature an analog of TTMs with which one could construct
strictly isospectral partners.  Moreover, by
(\ref{eq:vw}) and (\ref{eq:wphi}) the nodal condition maps
a regular $V$ to ${\tilde V} \sim 2/x^2$ for $x \rightarrow 0^+$
(generalizing the well-known result that SUSY increases the
angular momentum by one unit in the hydrogen atom).
It would therefore be interesting to see if an enlarged
class of transformations can address this question.

\subsection{Total-transmission modes}
\label{subsect:ttm}

The present paper refers in the main to states in $\Gamma$, i.e.,
states that satisfy the OWC at both extremes.  One could also
develop the same formalism for TTMs; see, e.g., the 
end of Section~\ref{sect:ex}.  Note that we here consider
TTMs as states on which SUSY acts, rather than
as ($\chi = 0$) generators.  To be more specific, 
the SUSY transformation acts on
the space $\Gamma_{\rm L}$ of TTM${}_{\rm L}$'s

\beq
\phi(x) \sim e^{i\omega x} \quad , 
\quad |x| \rightarrow \infty \quad ,
\eeql{eq:gaml}

\nid
or (equivalently, under $\omega \mapsto -\omega$)
the space $\Gamma_{\rm R}$ of TTM${}_{\rm R}$'s

\beq
\phi(x) \sim e^{-i\omega x} \quad , 
\quad |x| \rightarrow \infty \quad .
\eeql{eq:gamr}

\nid
One significant difference is that a $\chi = \pm 1$
transformation preserves the eigenvalues in these spaces,
whereas a $\chi = 0$ transformation may shift one
state.

\subsection{Tails}
\label{subsect:tail}

So far we have only considered potentials with
finite support.  We end with an outline
of some issues that arise for long-range potentials.

The most obvious difference is that if $V$ and/or ${\tilde V}$
are allowed to have tails that decay 
exponentially or slower in $|x|$, then the mixed-type SUSY
transformation is allowed, and there is a {\em continuous\/}
choice of generators.

However, there are a range of more subtle issues.  First,
the very definition of an outgoing wave requires care.
At the numerical level, 
special treatment is needed to ensure convergence\cite{tam},
but there are matters of principle as well
when the OWC $\phi(\omega, x) \sim e^{+i\omega |x|}$ is 
imposed only as $|x| \rightarrow \infty$.  In that limit, the
condition as stated becomes vacuous for $\mbox{Im } \omega < 0$,
since admixture of an (exponentially smaller) incoming solution
would not alter this behavior.  Rather, it is necessary to define the OWC 
in the upper half-plane in $\omega$ and analytically
continue to the lower half-plane.  A wave $\phi(\omega,x)$
is incoming if $\phi(-\omega,x)$ is outgoing~\cite{conv};
this is equivalent to saying that incoming waves are defined first
in the lower and continued to the upper half-plane.  
The necessity for these procedures
makes it possible that (at certain singular points $\omega$)
a wavefunction can be {\em both\/} incoming and outgoing.

Singularities in the one-sided functions~\cite{normfg}
can only occur because of the need to integrate the defining
equation over an infinite range.  Those values of $\omega$
for which these functions (say the left function $f$) are
singular are said to be {\em anomalous}; if the potential
is not oscillating, anomalous points can only occur on the
imaginary axis --- precisely where possible SUSY generators
are to be found.  The case of an exponential tail
$V(x) \sim V_1 e^{-\lambda |x|}$ is of particular interest
because the anomalous points
(at $\omega_m = -im\lambda/2$, $m = 1, 2, \ldots$)
can be studied by the Born approximation,
which in this case is equivalent to a power-series expansion
in $z = e^{-\lambda |x|}$.  More generally, if 

\beq
V(x) = \sum_{k=1}^{\infty} V_k e^{-k\lambda |x|}
\quad ,
\eeql{eq:genexp}

\nid
then for particular choices of the coefficients $V_k$,
one (or more) of the generically anomalous points $\omega_m$
may turn out to be regular
--- a situation we refer to as {\em miraculous\/}.
The anomalous and miraculous properties for the one-sided
functions $f$ and $g$ are inherited by their Wronskian $J$,
which is central to the formalism.
These concepts have been 
discussed in relation to a particular application~\cite{amb}.

As far as SUSY is concerned, we only make one 
remark: such singularities lead to poles
in $J(\omega)$, thus are associated with negative values
of $n(\omega)$, and are related in SUSY partners
by (\ref{eq:index3}).  The P\"oschl--Teller
potential $V(x) = {\cal V} \sech^2(x/b)$~\cite{pt} 
illustrates many interesting features,
including the existence of double poles of $J(\omega)$ which can merge
with two zeros and lead to ``nothing" --- a potential that
has total transmission at {\em all\/} energies~\cite{total}.

Anomalous points are exceptional
(in the case of exponential tails being a discrete
set of measure zero in the $\omega$-plane) and 
miraculous cancellation of singularities doubly exceptional.
One might therefore think that these are not important.
Surprisingly however, the problem of linearized
gravitational waves propagating on a black-hole background~\cite{bh}
is precisely miraculous in this sense, at the so-called
algebraically special frequency $\Omega$~\cite{alg}.
A generator at $\Omega$ leads to a SUSY transformation
that exactly relates the axial and polar sectors,
which are therefore isospectral in $\Gamma$.  
Among the more intriguing results is the following\cite{amb}: 
in the polar sector there is a mode at $\Omega$
that is simultaneously a QNM {\em and\/} a TTM [i.e., at radial infinity
it is purely outgoing but into the event horizon it is both outgoing
and incoming], while no modes exist in the axial sector.

The subtle and perhaps counter-intuitive nature of these concepts demands 
a separate and rigorous
examination, to which the foregoing is meant only as a preview
--- and as further illustration of the utility
of SUSY in dealing with waves in open systems.


\acknowledgments
The work is supported in part
by a grant (CUHK 4006/98P) from the Hong Kong Research Grants 
Council.  We thank Y.~T.~Liu,
C.~P.~Sun, B.~Y.~Tong and Jianzu Zhang for discussions.


\appendix



%

\section{Nodes in QNM wavefunctions}
\label{sect:appnode}

Nodeless eigenstates play a special role in SUSY:
they are candidates for the generator $\Phi$.
It is therefore useful to highlight
the differences between NMs and QNMs in this
regard, especially to contrast with the well-known
property that there can be {\em only one\/}
nodeless NM.

First of all, we show that for QNMs with 
$\mbox{Re } \omega \ne 0$,
there can be at most one node or antinode.  This is
not surprising: since the eigenvalue is complex, the
wavefunction has a changing phase, and it would be
``unlikely" that the real and imaginary parts
(or their derivatives) would vanish together.  To
prove this formally, take the Schr\"odinger point of view,
so that the eigenvalue is $E = \omega^2$ with
a nonzero imaginary part.  Now consider a time-dependent
QNM and suppose that it has nodes or antinodes at two
points $x_1, x_2$.  At these two points, the
current

\beq
{\cal J} = i \left[ \phi^* (\partial_x \phi) - (\partial_x \phi^*) \phi \right]
\eeql{eq:cons2}

\nid
vanishes.  Then, flux conservation implies that the total probability in the
interval $[x_1, x_2]$ is constant in time.  
Yet the wavefunction is either growing or decaying, 
since $\mbox{Im } E \ne 0$,
which is therefore a contradiction.

From the perspective of SUSY it is unfortunate that the above proof excludes
the crucial imaginary axis.  However, on that axis the statement remains
valid for repulsive potentials, or more generally for potentials which are
so weakly attractive that $V-\omega^2$ is positive definite~\cite{notpos}. 
Namely, let $\phi$ be a solution with two (anti)nodes. 
By taking the real or imaginary part, we may assume
$\phi$ to be real.  Now between two nodes
$\phi$ would have an extremum, i.e.\ $\phi''\phi<0$ which is incompatible
with the KGE. Similarly, an antinode can only be a global maximum or
minimum, precluding the presence of any other nodes or antinodes.

Thus, except for the imaginary axis in the case of attractive
potentials, QNMs can
have at most one node or antinode.  For symmetric potentials, in the even
sector $x=0$ is already an antinode, so there can be no nodes anywhere.

For zero modes, i.e., QNMs with $\mbox{Re } \omega = 0$,
nodes are more ``likely": the eigenvalue is
real and the wavefunction has a constant phase (say purely
real), so a node requires only {\em one\/} condition,
rather than two.  Nevertheless, in contrast
to the conservative case, the proof that 
there can be only one nodeless eigenstate can be bypassed.

The interlacing nodal structure of NM eigenfunctions
follows from well-established Sturm--Liouville 
theory.  For the present purpose, we do not need 
the full apparatus.  Consider, for simplicity,
a finite interval $[-a,a]$ and suppose there are two distinct
nodeless eigenfunctions $\phi_1 , \phi_2$,
both real.  Then they can both
be chosen to be positive, which violates the
orthogonality condition for NMs

\beq
\int_{-a}^{a} \phi_1(x) \phi_2(x) \, dx = 0 \quad .
\eeql{eq:orth1}

We can attempt to transplant the argument to QNMs.  For
zero modes, the wavefunctions can again 
be chosen to be real, and if they are nodeless,
positive definite.  However, the analog of
(\ref{eq:orth1}) for two eigenfunctions with eigenvalues
$\omega_j = -i\gamma_j$ is

\bea
&& -\left( \gamma_1 + \gamma_2 \right )
\int_{-a}^{a} \phi_1(x) \phi_2(x) \, dx
\nonumber \\
&&{} + \left[ \phi_1(-a) \phi_2(-a) + \phi_1(a) \phi_2(a) \right]
= 0
\quad .
\eeal{eq:orth2}

\nid
Note in particular the signs of the two terms.  With $\gamma_j > 0$,
this condition does {\em not\/} preclude both eigenfunctions
from being positive definite.

Thus, we can make three remarks.
(a)  For NMs, there can be {\em only one\/} nodeless state.
(b)  For QNMs with $\mbox{Re } \omega = 0$, there
{\em could be\/} more than one state with no node.
(c)  For QNMs with $\mbox{Re } \omega \ne 0$, or
with $\mbox{Re } \omega = 0$ but $V-\omega^2$ positive
definite, each
eigenfunction can have at most one node or antinode,
and for symmetric potentials, {\em every\/} even eigenfunction
is nodeless.

Case (b) in particular opens up the possibility of
multiple SUSY transformations.



%





%
%


\newpage
\leavevmode\scalefig{0.5}\epsfbox{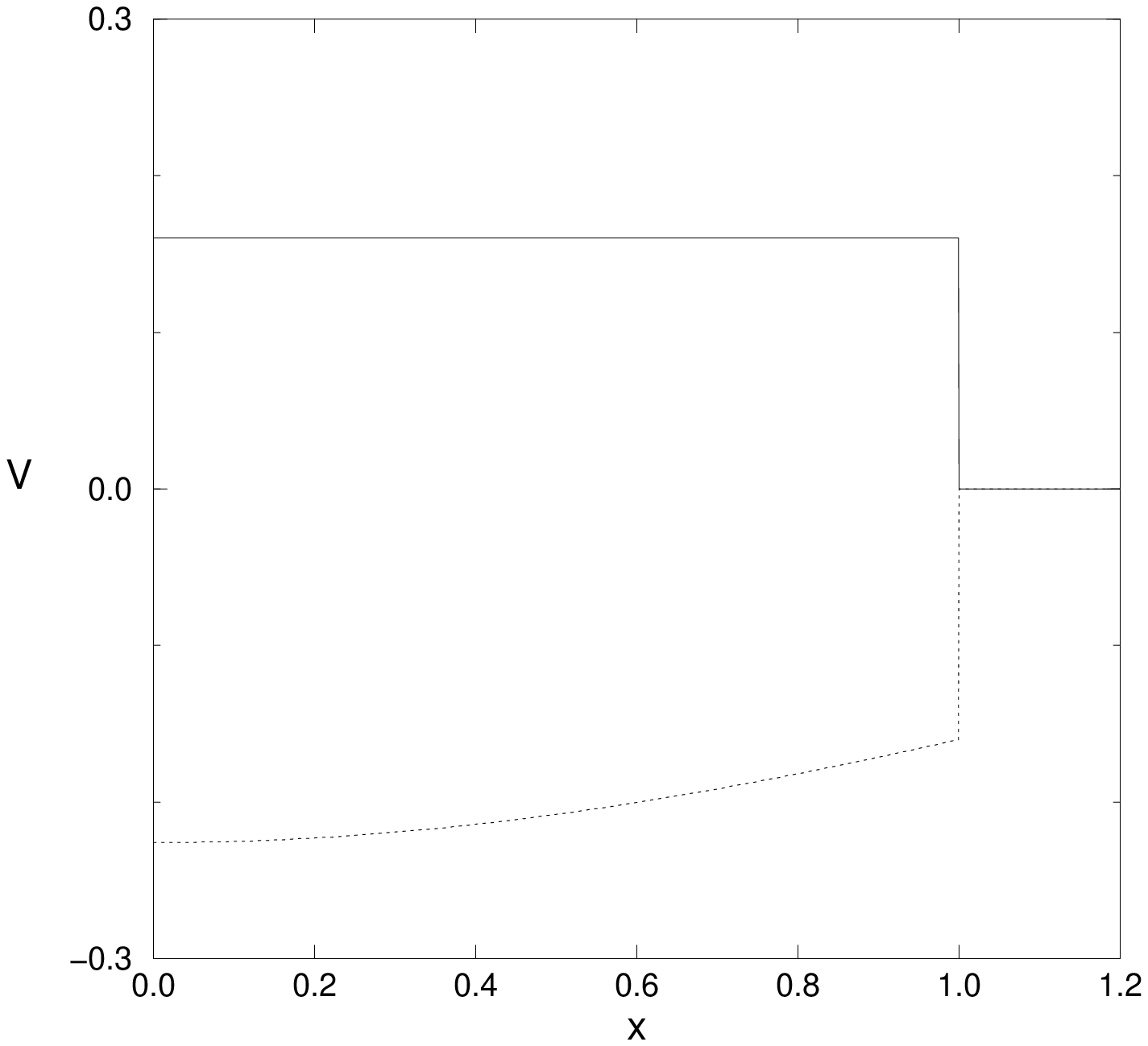}
\\
{\scriptsize Fig. 1(a): A square-barrier potential $V$ (solid line)
and its SUSY partner potential ${\tilde V}$ (broken line). 
Both potentials are symmetric and only the $x>0$ part is shown.
The SUSY transformation is constructed by using the state at 
$\Omega = \omega_1 = -0.181i$ [circle in Fig.~1(b)] as the generator.
}



\leavevmode\scalefig{0.5}\epsfbox{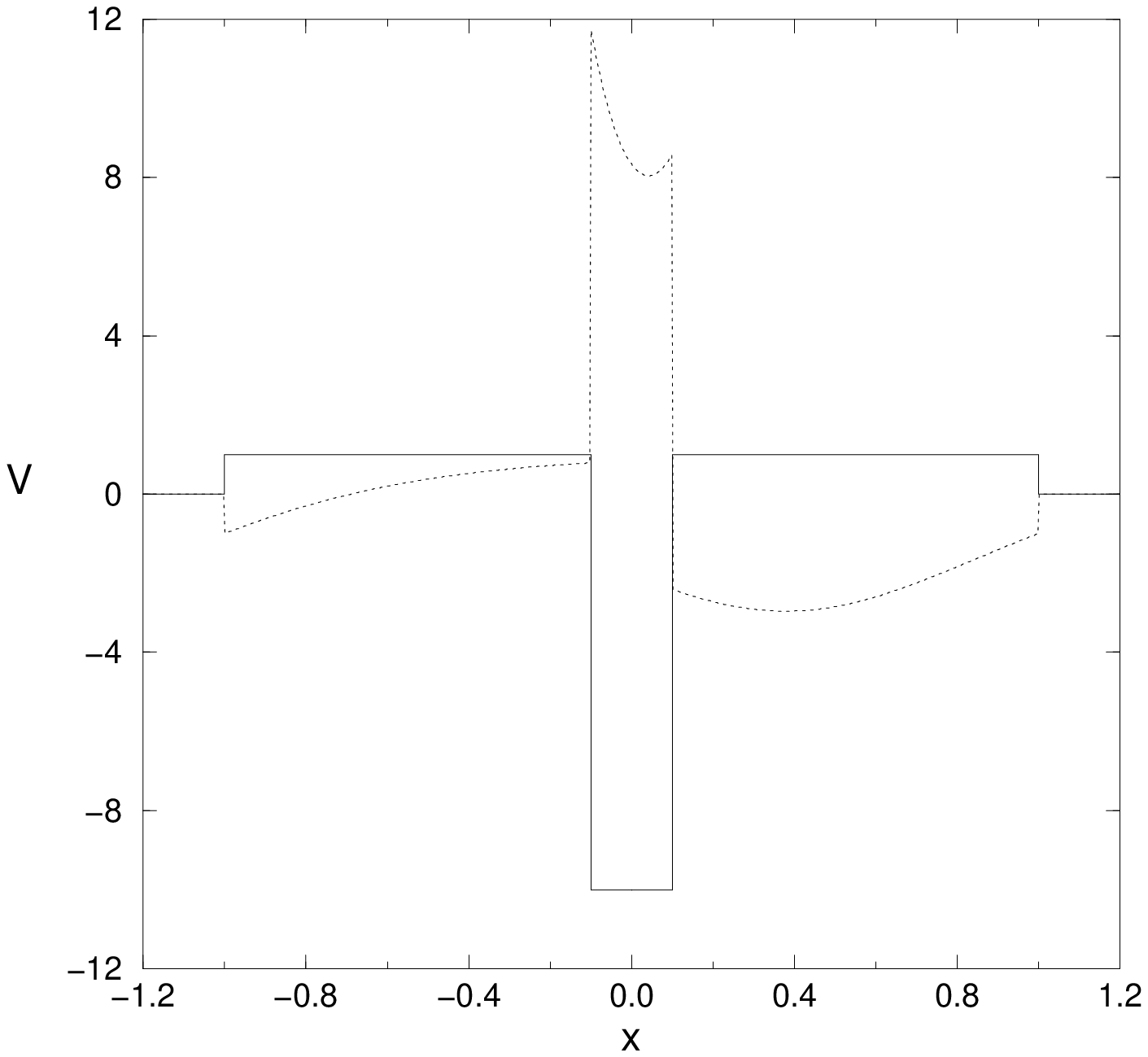}
\\
{\scriptsize Fig. 2(a):
A multi-step potential $V$ (solid line) 
and its SUSY partner potential ${\tilde V}$ (broken line).
The SUSY transformation is constructed by using the 
TTM$_{\rm L}$ at $\Omega = -0.990 i$ 
[square in Fig.~2(b)] as the generator. }



\leavevmode\scalefig{0.5}\epsfbox{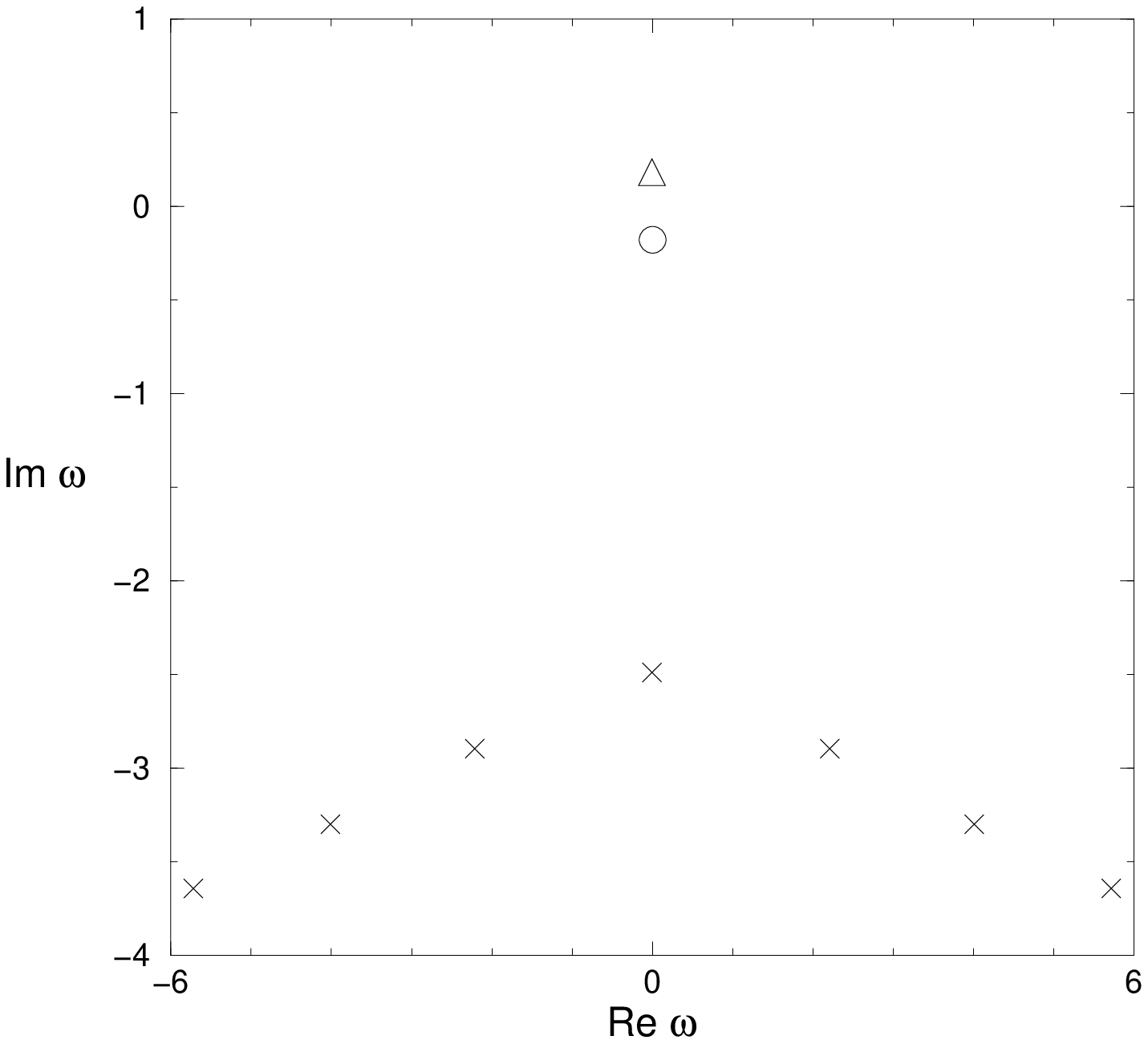}
\\
{\scriptsize Fig. 1(b): The complex $\omega$-plane showing the
QNMs common to both potentials (crosses); the mode present only in $V$
(circle), which corresponds to the generator $\Phi$; and the mode
present only in ${\tilde V}$ (triangle), which corresponds to ${\tilde
\Phi} = \Phi^{-1}$. }

\vspace{2cm}


\leavevmode\scalefig{0.5}\epsfbox{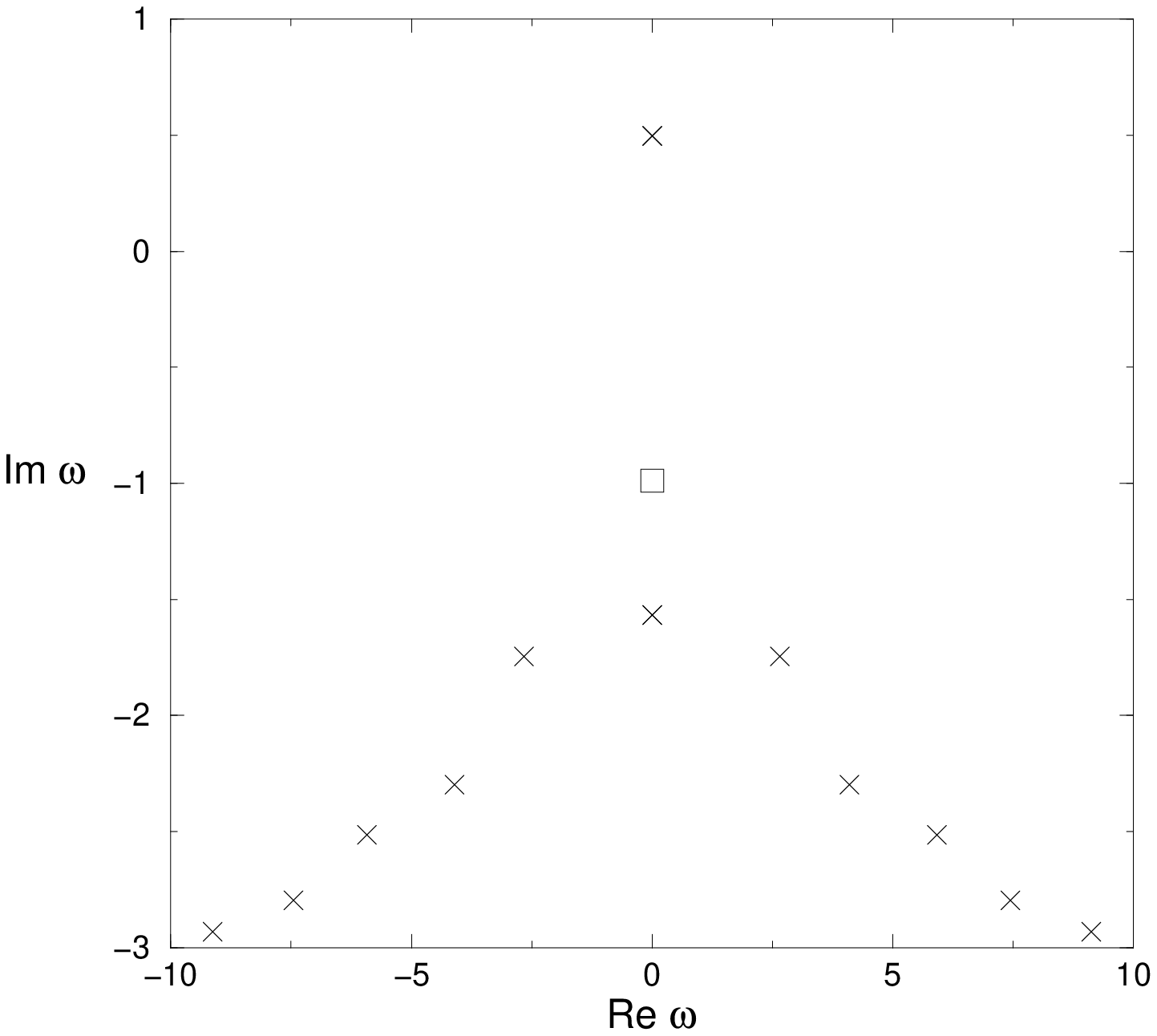}
\\
{\scriptsize Fig. 2(b):
The complex $\omega$-plane showing 
the NM and QNMs common to both potentials
(crosses). The square indicates a TTM${}_{\rm L}$
and a TTM${}_{\rm R}$ in $V$, and a doubled TTM${}_{\rm R}$ 
in ${\tilde V}$.
}

\end{multicols}


%

\newpage
\centerline{ {\bf FIGURE  CAPTIONS} }

\begin{description}

\item[Fig.~1] \mbox{ }\\
(a)  A square-barrier potential $V$ (solid line) 
and its SUSY partner potential ${\tilde V}$ (broken line).
Both potentials are symmetric and only the $x>0$ part is shown.
The SUSY transformation is constructed by using
the state at $\Omega = \omega_1 = -0.181i$ [circle in Fig.~1(b)]
as the generator.\\
(b)  The complex $\omega$-plane showing 
the QNMs common to both potentials
(crosses); the mode present only in $V$ (circle),
which corresponds to the generator $\Phi$; and
the mode present only in ${\tilde V}$ (triangle), which
corresponds to ${\tilde \Phi} = \Phi^{-1}$. 

\item[Fig.~2] \mbox{ }\\
(a)  A multi-step potential $V$ (solid line) 
and its SUSY partner potential ${\tilde V}$ (broken line).
The SUSY transformation is constructed by using 
the TTM$_{\rm L}$ at $\Omega = -0.990 i$ 
[square in Fig.~2(b)] as the generator. \\
(b) The complex $\omega$-plane showing the NM and QNMs common to both potentials (crosses). 
The square indicates a TTM${}_{\rm L}$ and a TTM${}_{\rm R}$ in $V$, and a doubled
TTM${}_{\rm R}$ in ${\tilde V}$. 

\end{description}




\begin{references}
\bibitem{conv}
The convention is that all time dependencies go as
$e^{-i\omega t}$, so $\omega$ and $-\omega$
in the time-independent function would describe time-reversed
states.
\bibitem{zeromode}
These are not to be confused with states with
zero eigenvalue, which are sometimes called
zero modes in the SUSY literature.  We shall not
use this latter nomenclature.
\bibitem{sumnm}
In the Fourier inversion of the Green's function
${\tilde G}(\omega)$~\cite{comp}, one has
to integrate along a line in the $\omega$-plane that
lies above all the singularities, to ensure that $G(t{<}0)=0$;
for $t{>}0$ this then captures all the singularities,
including the NMs.
\bibitem{oneqnm}
For a one-dimensional system with $V(|x|{>}a)=0$, starting at one end
(say $x = -a$) and using the known
logarithmic derivative $\phi'/\phi$ appropriate to
the OWC and an arbitrary
normalization, one gets a unique wavefunction.  Thus
there can be at most one eigenstate in $\Gamma$ at any $\omega$; there can
be no degeneracies.
\bibitem{missing}
If the double pole is split along the imaginary axis, 
there are two distinct SUSY transforms generated by 
$\tilde{\Phi}$ and $\tilde{\Psi}_j$ respectively, which become 
identical when the poles merge. 
Thus, in the spirit of the JB approach to the field expansion, 
one might try to find the ``missing" SUSY map generated 
by $\tilde{\Psi}_{j,1}$. However, no meaningful transformation emerges. 
This difference with the case of the field expansion can  
be understood by realizing that, when the poles split perpendicular 
to the imaginary axis upon changing the sign of the splitting 
perturbation, the SUSY transforms cease to exist altogether.
\bibitem{notunique}
The pre-image is unique only up to adding a multiple of
$f(\Omega,x)$; but this does not alter the conclusion that
none of these pre-images are in $\Gamma$.
\bibitem{normfg}
In this case the normalization convention is that
$f \sim 1 \cdot e^{-i\omega x}$ as $x \rightarrow -\infty$
and $g \sim 1 \cdot e^{i\omega x}$ as $x \rightarrow \infty$.
\bibitem{notpos}
Note that for an NM, $V-\omega^2$ cannot be positive definite.
\end{references}
\end{document}